\documentclass{article}
\usepackage{graphicx}
\usepackage{amssymb}

\def\0{\over } \def\1{\vec } \def\2{{1\over2}} \def\4{{1\over4}}
\def\6{\partial } \def\8#1{\textstyle #1}

\def\be{\begin{equation}}
\def\ee{\end{equation}}
\def\bea{\begin{eqnarray}}
\def\eea{\end{eqnarray}}
\def\bp{\begin{picture}(0,0)}
\def\ep{\end{picture}}
\def\nn{\nonumber}
\def\Tr{{\rm Tr }}
\def\tr{{\rm tr }}
\def\Im{{\rm Im\,}}
\def\Re{{\rm Re\,}}

\begin{document}
\setcounter{footnote}{0}
\centerline{\hfill\protect\raisebox{10pt}{\tt TUW-01-30}}
\renewcommand{\thefootnote}{\fnsymbol{footnote}}
\begin{flushleft}\large\bf
{HARD THERMAL LOOPS AND\\
QCD THERMODYNAMICS\footnote{Lectures given at the NATO Advanced Study
Institute ``QCD perspectives on hot and dense matter'', August 6--18, 2001,
in Carg\`ese, Corsica, France} }
\end{flushleft}
\vspace{5pt}
\begin{quote}
{\large\sc A. Rebhan}\\[2pt]
{\small Institut f\"ur Theoretische Physik\\[-1pt]
            Technische Universit\"at Wien\\[-1pt]
            Wiedner Hauptstr. 8-10, A-1040 Vienna, Austria}
\end{quote}
\vspace{5pt}
\begin{quote}\small
{\bf Abstract:}
These lectures give an introduction to thermal perturbation
theory, hard thermal loops, and their use in a nonperturbative, approximately
self-consistent resummation of the thermodynamical potentials of quantum
chromodynamics.
\end{quote}

\vspace{5pt}

\renewcommand{\thefootnote}{\arabic{footnote}}
\setcounter{footnote}{0}
\section{Introduction}

At sufficiently high temperature and/or density, quantum
chromodynamics\break (QCD) should become accessible by perturbative
methods due to asymptotic freedom. Thermal perturbation theory
is however a surprisingly intricate subject, and it was
only during the last decade that the necessary methodology
has been developed, and this development is not yet completely
finished. An important milestone was the hard-thermal-loop (HTL)
resummation programme introduced in particular by Braaten and
Pisarski \cite{Braaten:1990mz} which generalized the
static ring resummations that were known to be required
in the calculation of thermo-{\em static} quantities since long
\cite{Gell-Mann:1957,Akhiezer:1960} to dynamic quantities
such as quasiparticle properties as well as production and loss rates.

There are, however, well-known limitations of a fundamental nature
which present an impenetrable barrier to perturbation theory
at a certain order of the coupling (depending on the
quantity under consideration), caused by the 
inherently nonperturbative chromo-magnetostatic
sector of nonabelian gauge theories 
\cite{Polyakov:1978vu,Linde:1980ts,Gross:1981br}.
This does not mean that thermal perturbation theory is
completely futile, though, but rather that at particular
points certain nonperturbative input is required in addition
\cite{Braaten:1995na}, for example from lattice calculations
of the inherently nonperturbative 3-d Yang-Mills theory
describing the self-interactions of chromo-magnetostatic
fields. There are in fact even more (less well-known)
limitations from collinear singularities that occur
in real-time quantities involving external light-like momenta 
such as the production rate of real photons from
a QCD plasma \cite{Baier:1994zb,Aurenche:1996sh},
which require improved resummations \cite{Flechsig:1996ju}
and/or nonperturbative input.

But even in those cases where thermal perturbation theory has
not yet run into one of those barriers, such as the landmark
three-loop calculation of the free energy of QCD to order
$\alpha_s^{5/2}$ by Arnold, Zhai, and others
\cite{Arnold:1995eb,Braaten:1996jr}, 
there are severe problems caused by extremely poor convergence
and strong renormalizaton scheme dependences which seem to
render quantitative predictions possible only beyond preposterously
high temperatures.
While some improvement seemed possible through tricks
like Pad\'e approximations 
\cite{Kastening:1997rg,Hatsuda:1997wf},
the sad conclusion appeared to be that thermal perturbation theory
was not applicable in QCD at temperatures of practical interest
\cite{Braaten:1996ju}.

Recently, however, it became clear that the problem of poor
convergence is not specific to QCD, but arises already
in such simple theories as massless scalar $\varphi^4$ theory
\cite{Drummond:1997cw} and that alternative resummations
can be found that greatly improve the apparent convergence
\cite{Karsch:1997gj,Andersen:2000yj}. In these lectures,
after an introduction to hard thermal loops, I will
present an approach to the problem of calculating
thermodynamical quantities in QCD in an {\em approximately
self-consistent} resummation of hard thermal loops
and their next-to-leading order corrections 
\cite{Blaizot:1999ip,Blaizot:1999ap,Blaizot:2000fc,Blaizot:2001vr}
that appears to work well down to temperatures a few times
the deconfinement transition temperature and that
suggests that at such temperatures
the still strongly interacting QCD may
(at least in certain cases) be adequately
described by weakly interacting HTL quasiparticles after all.

\section{Thermal field theory}

The Feynman rules of thermal field theory \cite{Kap:FTFT,LeB:TFT}
are most easily
formulated in the imaginary-time (ITF) or Matsubara formalism.
The statistical density operator $e^{-\beta H}$ is then
equivalent to a time evolution operator over an imaginary
time interval of length $\beta$, the inverse temperature $T^{-1}$,
and the traces in $\langle A \rangle = {\rm Tr}[e^{-\beta H} A]$
require (anti-)periodic boundary conditions at the ends
of the imaginary time interval for bosons (fermions).
This gives rise to discrete imaginary (Matsubara) frequencies
when going to momentum space, so that the only change
in the Feynman rules is in the replacement
\be
\int {d^4k\over i(2\pi)^4} \to 
\beta^{-1}\sum_\nu \int {d^3k\over (2\pi)^3},\qquad
i(2\pi)^4 \delta^4(k) \to\beta (2\pi)^3 \delta_{\nu,0} \delta^3(k).
\ee
with
\be
k_0\to 2\pi i T \nu,\qquad 
\nu\in \left\{ \begin{array}{ll}\mathbb Z & \mbox {bosons} \\
                              \mathbb Z -{1\over2} &  \mbox {fermions} \\
               \end{array} \right. \;.
\ee

Since it is usually hard to evaluate the resulting sums directly,
they are best turned into integrals again by writing, in the bosonic case,
\bea\label{freqsum}
&&T\sum_{n=-\infty}^\infty f(k_0=2\pi i n T)=
{T\02\pi i}\oint_{\cal C} dk_0 f(k_0){\beta\02}{\rm coth}{\beta k_0\02} \nn\\
&=&\underbrace{ {1\02\pi i} \int_{-i\infty}^{i\infty}f(k_0) }_{\mbox{vacuum
contribution}}+\underbrace{ {1\02\pi i} 
\int_{-i\infty+\epsilon}^{i\infty+\epsilon} 
\left[f(k_0)+f(-k_0)\right]{1\0e^{\beta k_0}-1} }_{\mbox{thermal contribution}}
\eea
for a function $f(k_0)$ that is regular for $k_0 \in i\mathbb R$ and
where $\cal C$ is a contour encircling only the poles of coth at the
location of the Matsubara frequencies. (In the fermionic case, an analogous
formula can be easily obtained by using tanh in place of coth, which
in the right-hand side
leads to minus the Fermi-Dirac distribution in place of the
Bose-Einstein one.)

This evaluation through contour integrals leads to a nice separation
into a (Wick rotated) $T=0$ contribution, and a purely thermal one,
which vanishes for $\beta\to\infty$, i.e. $T\to0$.
There exists also a formulation directly in real time (Schwinger-Keldysh
and variants thereof) \cite{Landsman:1987uw}
where this separation is already conspicuous
in the Feynman rules. However, this requires a doubling of fields,
a $2\times2$ matrix structure for propagators and even more
components for $n$-point vertex functions, which in fact
correspond to the many possibilities of analytic continuation to
real frequencies (in particular if there are several independent
external frequencies) \cite{Kobes:1990kr}. 

A simple case that leads to a thermal contribution in (\ref{freqsum})
is the frequently occurring one that there is a simple pole
in $f(k_0)$ at $k_0=\pm E$ with $E>0$. Then one has
\be
T \sum_n {1\0 k_0 \pm E} = \pm (-1)^\sigma n_\sigma(E) 
\,+\, \mbox{vacuum contributions},
\ee
where $n_\sigma(E)=[e^{\beta E}-\sigma]^{-1}$ and $\sigma=+1$ for
bosons, $-1$ for fermions.

\section{Hard thermal loops}

The simplest example of a hard thermal loop is given by the
one-loop self-energy diagram in a scalar $g\varphi^4$ theory.
This is a tadpole diagram, independent of external momenta:
\be
\Pi={4! g^2\02} \sum_K {-1\0K^2-m^2}
\ee
where we have introduced the notation
$\sum_K = T \sum_n \int{d^3k\0(2\pi)^3}$,
$K^\mu = (k^0,k^m)$.

With $E=\sqrt{k^2+m^2}$, its thermal contribution is easily
evaluated as
\bea
\sum_K {-1\0K^2-m^2}&=&
\sum_K {1\02E}\left( {1\0k_0-E} - {1\0k_0+E} \right) \nn\\&=&
\int{d^3k\0(2\pi)^3} {1\02E} 2n(E)
= \int_0^\infty {dk\,k^2\02\pi^2 E}n(E). 
\eea
(The vacuum contribution is removed by standard ($T$=0)
renormalization.)
Without the Bose-Einstein factor, this integral would be
quadratically divergent. Thanks to the former, it is finite,
but dominated by momenta $k \sim T$. If $T\gg m$, 
the mass terms of the $T$=0 theory can be neglected, and
the leading contribution to the self-energy is given
by a self-energy contribution proportional to $T^2$,
which is called a {\em hard thermal loop} (HTL):
\be
\hat\Pi = {4! g^2\02} \sum_K {-1\0K^2}=
g^2{6\0\pi^2}\int_0^\infty dk\,k\,n(k)= g^2T^2 \;,
\ee
where the hat is a reminder of the HTL approximation -- in this
case the neglect of the bare mass $m$.

$\hat \Pi$ clearly corresponds to a mass term for scalar excitations
generated by interactions with the particles in the heat bath.
It should be understood, however, that this {\em thermal mass}
is qualitatively different from ordinary masses. In particular,
one can show that unlike ordinary masses
it does not spoil conformal invariance (if any) as it
does not contribute to the trace of the
energy momentum tensor \cite{Nachbagauer:1996wn}.

This example for a HTL is, however, deceptively simple.
In general, thermal masses are not constant but depend
on momentum, that is, they correspond to nonlocal terms in
an effective action.

The thermal masses generated for gauge fields are of this
form. Indeed, a constant mass term would violate gauge invariance.
Let us consider as a simple gauge theory example the case
of scalar electrodynamics with covariant gauge fixing,
\be
{\cal L}=(D_\mu \phi)^* D^\mu \phi -{1\04} F_{\mu\nu} F^{\mu\nu}
-{1\02\alpha}(\6_\mu A^\mu)^2
\ee
where $D_\mu=\6_\mu+ieA_\mu$. The tree-level propagator
reads $G_{\mu\nu}^0=g_{\mu\nu}/K^2+(\alpha-1)K_\mu K_\nu/K^4$
and this will receive thermal corrections through the
photon polarization tensor, which in momentum space reads
\be\label{PiSED}
\Pi_{\mu\nu}(K)=e^2 \sum_P \left[
{(2P-K)_\mu (2P-K)_\nu \0 P^2 (P-K)^2} - {2g_{\mu\nu}\0P^2} \right].
\ee
The last term is from the seagull diagram, which is essentially
the same as the tadpole diagram before. The first term is however
dependent on the external momentum, and there is no reason to
expect a Lorentz invariant form, because the heat bath is
singling out a preferred frame of reference.

There are in fact four symmetric tensors that can be built from
the available quantities $g_{\mu\nu}$, $K_\mu$, and the
four-velocity of the heat bath which we have
tacitly chosen as $U_\mu=\delta_\mu^0$. 

In electrodynamics, the polarization tensor has to be transverse,
$K^\mu \Pi_{\mu\nu}$ $\equiv0$. This additional requirement still
leaves two possible tensors, for from $U$ and $K$ one can
build the transverse vector $\tilde U_\mu= K^2 U_\mu - (U\cdot K)K_\mu$.
We choose them as
\be
A_{\mu\nu}=g_{\mu\nu}-{K_\mu K_\nu/K^2}-B_{\mu\nu},\quad
B_{\mu\nu}=\tilde U_\mu \tilde U_\nu / \tilde U^2.
\ee
$A_{\mu\nu}$ can easily be shown to have vanishing components
$A_{0\nu}=0=A_{\mu0}$, whereas $A_{ij}=-\delta_{ij}+k_i k_j/k^2$,
so this is a projection onto spatially transverse momenta;
$B_{\mu\nu}$ is a projector orthogonal to $A_{\mu\nu}$.

In nonabelian gauge theories like QCD, one generally has a more
complicated structure. Then one needs two more, nontransverse, tensors,
which one may choose as
\be
C_{\mu\nu}={\tilde U_\mu K_\nu + K_\mu \tilde U_\nu \0\sqrt2 K^2 k},
\quad D_{\mu\nu}={K_\mu K_\nu\0K^2}\;.
\ee

Decomposing $\Pi_{\mu\nu}=-\Pi_t A_{\mu\nu}- \Pi_\ell B_{\mu\nu}-
\Pi_c C_{\mu\nu}-\Pi_d D_{\mu\nu}$ and
\be
G^{\mu\nu}\equiv [G_0^{-1}+\Pi]^{-1\mu\nu}=
\Delta_t A_{\mu\nu} +\Delta_\ell B_{\mu\nu}+
\Delta_c C_{\mu\nu}+\Delta_d D_{\mu\nu}
\ee
one has
\bea
\Delta_t &=& [K^2-\Pi_t]^{-1}, \quad
\Delta_\ell = [K^2-\Pi_\ell+\alpha{\Pi_c^2\0K^2-\alpha \Pi_d}]^{-1}, \\
\Delta_c&=& \alpha \Pi_c [K^2-\alpha\Pi_d]^{-1},\quad
\Delta_d = {\alpha (K^2-\Pi_\ell)\0K^2-\alpha \Pi_d}\Delta_\ell\;.
\eea

In view of the explicit gauge parameter dependences (there are
in fact more hidden within the structure functions of $\Pi$), it is
remarkable that one can prove that the singularities
of $\Delta_t$ and $\Delta_\ell$ are gauge-fixing independent
\cite{Kobes:1990xf,Kobes:1991dc,Rebhan:2001wt}.

In electrodynamics this situation is much simpler (unless one
introduces nonlinear gauge fixing): $\Pi_{\mu\nu}$ is both
transverse and completely gauge parameter independent.

Because of transversality, which is easily verified for
(\ref{PiSED}), there are only two independent components,
e.g.\ $\Pi^\mu_{\;\mu}$ and $\Pi_{00}$.
The former reads
\be\label{PimumuSED}
\Pi^\mu_{\;\mu}(K)=e^2 \sum_P\left[ {-4\0P^2} - {K^2\0P^2(P-K)^2} 
\right] = 4e^2 \sum_P{-1\0P^2} + O(T) = {e^2T^2\03} + O(T)
\ee
where the term proportional to $K^2$ does not constitute
a hard thermal loop, because its integrand does not involve
a quadratic divergence in its vacuum piece. For $k_0,k \ll T$,
$\Pi^\mu_{\;\mu}(K) \sim \hat\Pi^\mu_{\;\mu}=e^2 T^2/3$.
Notice, however, that this `HTL approximation' remains valid
even for $k_0, k\sim T$ as long as $K^2 \ll T^2$.

The other component, $\Pi_{00}$ is more complicated to evaluate.
Its HTL piece, which is contained in
\be\label{Pi00SED}
\Pi_{00}(K)=e^2\sum_P \left[ {4p_0 (p_0-k_0)\0P^2 (P-K)^2}-{2\0P^2}
\right],
\ee
can however be extracted rather easily using
$$
{p_0\0P^2}=\2({1\0p_0-p}+{1\0p_0+p}),\quad
{1\0p_0+X}{1\0p_0+Y}={1\0X-Y}({1\0p_0+Y}-{1\0p_0+X})
$$
which gives a number of terms of the form ${1\0X-Y}(n(Y)-n(X))$
after summing over the Matsubara frequencies.
Now, HTL contributions arise from $p\sim T$, and for
$k_0,k \ll T$ one can approximate the energies $X$ and $Y$ by $\pm p$,
except when two hard energies form a soft difference. This
gives
\bea
\Pi_{00}(K)\sim e^2 \int{d^3p\0(2\pi)^3} \biggl\{\!\!\!\!\!\!\!\!\!\!\!&&
{1\0k_0+p-|p-k|}[n(|p-k|)-n(p)] \nn\\
&+&{1\0k_0-p+|p-k|}[-n(|p-k|)+n(p)]\, \biggr\}
\eea
where the tadpole-like last term in (\ref{Pi00SED}) has cancelled
against terms where the energy denominators contain the sum of
two hard energies.

Since $k\ll p\sim T$, we have $n(|p-k|)-n(p)\sim n'(p)[|p-k|-p]$
and $|p-k|-p \sim \1p\cdot\1k/p\equiv zk$, so the HTL piece of $\Pi_{00}$ is
finally given by
\bea\label{Pi00}
\hat\Pi_{00}(K)&=&2e^2 \int {p^2\,dp\0(2\pi)^2}n'(p)
\int_{-1}^1 dz \left[ -1 + {k_0\0k_0-zk} \right] \nn\\
&=& {e^2T^2\03} \left[ 1 - {k_0\02k} \ln {k_0+k \0 k_0-k} \right].
\eea

Originally, $k_0$ was restricted to Matsubara frequencies. In order
to allow for soft $k_0\ll T$ without being restricted to the
zero mode, we in fact need analytic continuation, 
e.g.~$k_0 \to \omega+i\epsilon$ for retarded boundary conditions,
and this defines the cut of the logarithm in (\ref{Pi00}).

The above results for $\hat\Pi^\mu_{\;\mu}$ and $\hat\Pi_{00}(K)$
are actually universal. They have the same form in nonabelian
gauge theories, only the overall coefficient differs. In SU($N$)
gauge theories with $N_f$ quark flavors one just needs to
replace \cite{Kalashnikov:1980cy,Weldon:1982aq}
$e^2 \to g^2(N+N_f/2)$. They also retain their form in the
presence of a nonvanishing chemical potential $\mu_f$, which
leads to the replacement of $T^2\to T^2+3\mu_f^2/\pi^2$ in
the contributions $\propto N_f$.
In the HTL approximation, there
are moreover no nontransverse contributions\footnote{At order $kT$
one has however $\Pi_c\not=0$ for $\alpha\not=1$ in the nonabelian case,
and $\Pi_d\not=0$ at two-loop order whenever $\Pi_c\not=0$.}
so the HTL gauge boson propagator involves two independent
branches determined by
$\hat\Pi_\ell = -\hat\Pi_{00} K^2/k^2$ and $\hat\Pi_t=(\hat\Pi^\mu_{\;\mu}
-\hat\Pi_\ell)/2$. 

\begin{figure}
\includegraphics[viewport = 0 160 540 550,scale=0.44]{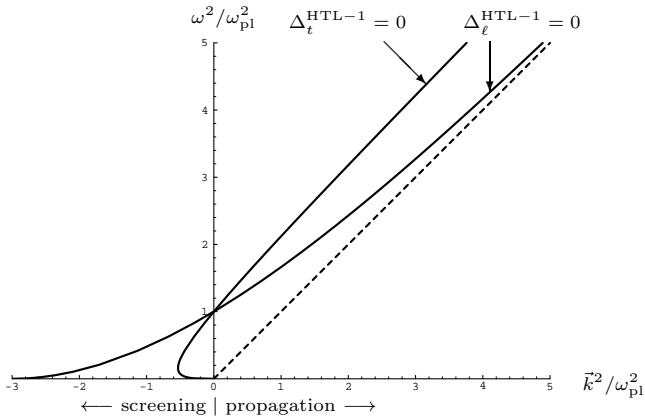} 

\bp\scriptsize
\put(218,18){$\vec k^2/\omega_{\rm pl}^2$}
\put(70,157){$\omega^2/\omega_{\rm pl}^2$}
\put(107,155){$\Delta_t^{{\rm HTL}-1}=0$}
\put(173,155){$\Delta_\ell^{{\rm HTL}-1}=0$}
\put(143,150){\vector(1,-1){16}}
\put(183,150){\vector(0,-1){19}}
\put(78,10){$|$ propagation $\longrightarrow$}
\put(28,10){$\longleftarrow$ screening}
\ep
\caption{The location of the zeros of $\Delta_t^{{\rm HTL}-1}$ (spatially
transverse gauge bosons) 
and of $\Delta_\ell^{{\rm HTL}-1}$ (longitudinal plasmons) in quadratic
scales such as to show propagating modes and screening phenomena
on one plot. 
}\label{Figg}
\end{figure}

The poles of the propagators $\Delta_t$ and $\Delta_\ell$
determine the dispersion laws of two sorts of quasiparticles,
which 
in contrast to the scalar $\varphi^4$ example are not given
by simple mass hyperboloids. These are displayed in Fig.~\ref{Figg}
in a plot of $\omega_{t,\ell}(k)$ in quadratic scales (where
a relativistic mass hyperboloid would show up as a straight line
parallel to the light-cone).

Above a common {plasma frequency} $\omega_{\rm pl.}=eT/3$,
there are propagating modes, which for large
momenta in the transverse branch
tend to a mass hyperboloid with asymptotic mass 
$m_\infty^2={3\over 2}\omega_{\rm pl.}^2$, and in branch $\ell$ approach the
light-cone exponentially with exponentially vanishing residue.
Indeed, this mode does not have an analogue in the $T$=0 theory
but is a purely collective phenomenon, so it has to disappear from
the spectrum as $k\to\infty$. The spatially transverse mode, on
the other hand, represents quasiparticles that are in-medium
versions of the physical polarisations of gauge bosons.

For $\omega<\omega_{\rm pl.}$, $|\mathbf k|$ is the inverse screening
length, which in the static limit vanishes for mode $t$ (absence of
magnetostatic screening), but reaches the {Debye mass},
$\hat m_D^2=3\omega_{\rm pl.}^2$, for mode $\ell$
(electrostatic screening).
A vanishing magnetic screening mass is required by gauge invariance
in abelian gauge theories \cite{Fradkin:1965,Blaizot:1995kg},
but not in the nonabelian case. In fact, lattice simulations
of gauge fixed propagators in nonabelian theories do find
a screening behaviour in the transverse sector, 
however the corresponding singularity is certainly quite
different from a simple pole \cite{Cucchieri:2001tw}.

For $\omega^2<k^2$, there is a large imaginary part $\sim e^2T^2$
from (\ref{Pi00})
which prevents the appearance of poles in this region. This
imaginary part corresponds to the possibility of Landau damping,
which is the transfer of energy from soft fields to
hard plasma constituents moving in phase with the field 
\cite{LifP:PK,Blaizot:2001nr}
and is an important part of the spectral density of HTL
propagators. At higher, subleading orders of perturbation theory, it
is, however, not protected against gauge dependences
in nonabelian gauge theories.

To complete the discussion of HTL's in scalar electrodynamics, 
let us also consider briefly the scalar self-energy. In contrast
to $\varphi^4$ theory, this is now a nonlocal quantity. Nevertheless,
the HTL part is still a constant thermal mass as given by the
first term in
\be\label{Xi}
\Xi={e^2T^2\04}+e^2 K^2 \sum_P
\left[ {3-\alpha \0 P^2 (P-K)^2}+{2(\alpha-1) K\cdot P\0P^4 (P-K)^2}
\right].
\ee 
The other terms are not proportional to $T^2$ because the integrands
do not grow sufficiently at large momenta. They are even gauge parameter
dependent, in contrast to the HTL piece. Notice that these gauge
dependent terms vanish on the lowest-order mass shell $K^2=0$.

In QCD (and already in spinor QED), we also need to consider
the fermion self-energies \cite{Klimov:1981ka,Weldon:1989ys}. 
In the ultrarelativistic high-temperature 
limit, bare masses can be neglected. In this case the fermion
self-energy can be paramet\-rized by
\be
\Sigma(K)=a(K) \gamma^0 + b(K) \1k\cdot\1\gamma /k \;.
\ee
Using the same methods as above one easily computes the
HTL contributions in
\bea
\4 {\rm tr} \not\!\! K \Sigma &=& \omega a+k b = {e^2 T^2\08}\equiv \hat M^2,\\
\4 {\rm tr} \gamma^0 \Sigma &=& a = {e^2 T^2\016k} \ln{\omega+k\0\omega-k},
\eea
where in nonabelian gauge theories now $e^2\to g^2(N^2-1)/(2N)$.

The structure of the HTL fermion propagator is
\be
S(K)=\2\left(\gamma^0+{\1k\cdot\1\gamma\0k}\right) \Delta_+
+\2\left(\gamma^0-{\1k\cdot\1\gamma\0k}\right) \Delta_-
\ee
with $\Delta_\pm^{-1}=-[\omega \mp (k + \Sigma_\pm)]$ and
$\Sigma_\pm\equiv b\pm a$. The two branches correspond to
spinors whose chirality is equal (+) or opposite $(-)$ to their
helicity.

\begin{figure}
\includegraphics[viewport =                           
0 120 540 570,scale=0.38]{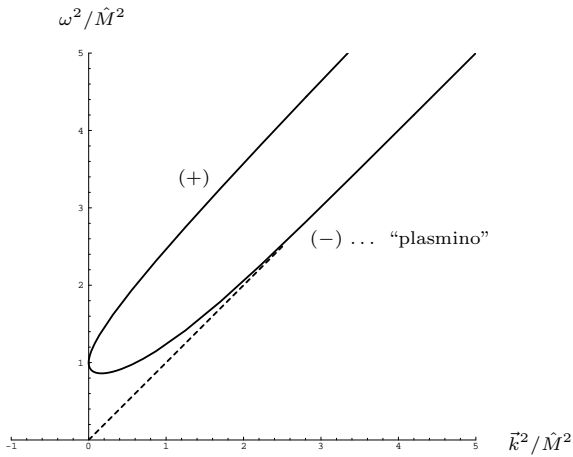} 

\bp\scriptsize
\put(190,14){$\vec k^2/\hat M^2$}
\put(20,175){$\omega^2/\hat M^2$}
\put(65,115){$(+)$}
\put(115,92){$(-)$ \ldots { ``plasmino''}}
\ep
\caption{The location of the zeros of $\Delta_\pm^{-1}$ in the
HTL approximation in quadratic scales.}
\label{Figf}
\end{figure}

The additional collective modes of branch $(-)$ (``plasminos'')
have a curious minimum of $\omega$ at $\omega/\hat M\approx 0.93$ and
$|\vec k|/\hat M\approx 0.41$ and approach the light-cone for
large momenta, but with exponentially vanishing residue.
The regular branch approaches a mass hyperboloid (in Fig.~\ref{Figf} a
straight line parallel to the diagonal) with asymptotic mass
$\sqrt2 \hat M$. Again, for space-like momenta, $K^2<0$,
there is a large imaginary part corresponding to Landau
damping, which now corresponds to the transmutation of
soft fermionic fields together with hard fermionic (bosonic) plasma
constituents into hard bosonic (fermionic) ones.

\section{Standard HTL perturbation theory}

When writing down dressed propagators, as we have done above
in the discussion of the spectrum of thermal quasiparticles,
we have already performed a resummation of infinitely many
loops---the geometric series of self-energy insertions according
to Dyson's equation, which in the scalar $\varphi^4$ theory
is simply
\be
{-1\0K^2-(gT)^2}={-1\0K^2}\left(1+g^2{T^2\0K^2}+g^4{T^4\0K^4}+\ldots
\right).
\ee
Clearly, the perturbative version is useful only when $K^2\gg (gT)^2$.
In particular, it fails for $k_0,k \sim gT$, the scale where
collective phenomena transform the familiar quanta into quasiparticles.

The appearance of thermal masses
presents unavoidable problems at some order of perturbation theory,
namely when these higher orders probe the energy/momentum scale
$k_0,k \sim gT$, and, in particular in the bosonic sector,
the Bose-Einstein distribution factors enhance the sensitivity
to the infrared.

It is therefore mandatory to switch from bare perturbation theory
to one that uses resummed (dressed) propagators. In simple cases
like scalar $\varphi^4$ theory, where the only HTL is a constant
mass term, this resummation has been studied
already long ago \cite{Dolan:1974qd}. 
Conceptually (although
not practically) equally simple is the example of scalar
electrodynamics, where the only HTL's are the self energy
diagrams considered above. However, already in spinor QED and
to a larger extent in QCD, it turns out that there are HTL vertex
functions \cite{Frenkel:1990br}
which have to be treated on a par with the HTL self
energies to achieve a systematic resummed perturbation theory
\cite{Braaten:1990mz}. Indeed, if $N$-point vertex functions give rise
to HTL's $\propto T^2$, they are as important as bare vertices
when the momentum scale is $\sim gT$:
\be
\quad { \Gamma_{,N}^{\rm HTL}} \sim g^N { T^2} { k^{2-N}}
\sim g^{N-2} { k^{4-N}} 
\sim {\partial ^N\mathcal L_{\rm cl.}\over \partial A^N}\Big|_{ k\sim gT}.
\ee
In fact, even for $N$ so high that there is no comparable tree-level vertex,
such vertex functions have to be resummed. Already in spinor
QED, there exist one-loop HTL vertex functions with two external
fermion lines and an arbitrary number of gauge boson lines.
In QCD, there are even more HTL vertex functions without
tree-level analogue. It turns
out that one-loop diagrams with an arbitrary number of external
gauge boson lines are HTL.

\subsection{HTL effective actions}

The HTL resummation programme can be understood as the transition
from the fundamental Lagrangian to an effective one generated
by `integrating out' the hard momentum modes $k_0,k\sim T$
in one-loop order.

In scalar $\varphi^4$ theory, this effective theory differs
from the bare one only in a simple thermal mass term,
$\mathcal L^{\rm HTL}_{\rm scalar}=-\2\mu^2\varphi^2$.

For gauge bosons and fermions, 
the effective theory is necessarily a non-local one
as gauge invariance forbids a simple local thermal mass term
and 
in the case of fermions it is due to the fact that
HTL's do not spoil chiral symmetries.
Remarkably, the infinitely many non-local vertex functions
can be summarized by a comparatively simple and manifestly gauge-invariant
integral representation 
\cite{Taylor:1990ia,Braaten:1992gm,Frenkel:1992ts}
\bea
\mathcal L^{\rm HTL} &=& \hat M^2 \int {d\Omega_v\over 4\pi} 
\bar \psi \gamma^\mu
{{ v_\mu}\over { v}\cdot D(A)} \psi \nonumber\\
&& -{3\over 2}\omega^2_{pl.} {\rm tr} \int {d\Omega_v\over 4\pi} F^{\mu\alpha}
{{ v_\alpha} { v^\beta} \over  ({ v}\cdot D_{adj.}(A))^2} F_{\mu\beta}
\label{HTLeffL}
\eea
$ v=(1,\vec v)$ is a light-like 4-vector, i.e.\ with $ \vec v^2=1$, 
and its
spatial components are averaged over by $ d\Omega_v$. 
$ v$ is the remnant of the hard plasma constituents'
momenta $p^\mu \sim T v^\mu$, namely their light-like
4-velocity, and the overall scale $T$ has 
combined with the coupling constant to form the scale of
thermal masses, $\hat M, \omega_{pl.}\sim gT$.

The covariant derivatives in the denominators of (\ref{HTLeffL}) 
are responsible for
the fact that there are infinitely many HTL's involving
external fermions and an arbitrary number of gauge bosons,
even in QED, where only the pure gauge-field sector becomes
bilinear because of $D_{adj.}(A)\to\partial$.

Technically, resummed perturbation theory amounts to
the replacement
\be\label{resum}
\mathcal L_{\rm cl.} \to \mathcal L_{\rm cl.} + \mathcal L^{\rm HTL} 
- \ell \mathcal L^{\rm HTL}
\ee
where $\ell$ is a loop counting parameter that is sent to 1 in the
end, after the last term has been treated as a `thermal counterterm'.

Because $\mathcal L^{\rm HTL}$ has been derived under the
assumption of soft external momenta, this prescription is in
fact only to be followed for soft propagators and vertices
\cite{Braaten:1990mz}.
Those involving hard momenta (if present) do not require this resummation,
and they can be excluded from this resummation by the introduction
of some intermediate scale $\Lambda$ with $gT \ll \Lambda \ll T$.
Complete results have to come out independent of $\Lambda$, of course.

\subsection{Example: NLO terms in scalar electrodynamics}

Massless scalar electrodynamics \cite{Kraemmer:1995az}
is a particularly simple toy model as
its HTL effective action (\ref{HTLeffL}) is bilinear in all fields.
The scalar HTL self-energy (\ref{Xi}) is moreover a simple mass term,
and in order to consider the one-loop corrections to the photon
polarization tensor, nothing more is needed. Let us consider two
limiting cases of this to illustrate some important points of
the HTL resummation programme.

\subsubsection{Debye mass}
\label{mDSED}

The Debye mass, i.e.~the inverse screening length of electrostatic
fields, is determined by the 
zero of $\Delta_\ell(k_0=0,k)$ at imaginary $k$
leading to
$m_D^2=\Pi_{00}(k_0=0,\1k^2=-m_D^2)$.
Because at leading order $\hat\Pi_{00}(k_0=0,k)$ turns out to
be independent of $k$ (see (\ref{Pi00})), the frequently
found definition \cite{Kap:FTFT} of $m_D^2$ as $\hat\Pi_{00}(k_0=0,k\to0)$
happens to be correct, but becomes unphysical in general \cite{Rebhan:1993az}:
beyond LO, only the former, self-consistent definition is
renormalization-group invariant and (in nonabelian theories)
gauge invariant.

The resummation programme sketched above makes the LO result
$\hat m_D^2=e^2T^2/3$ part of the new lowest-order Lagrangian
$\mathcal L_{\rm cl.} + \mathcal L^{\rm HTL}$.
The NLO correction therefore is given by one-loop diagrams
using this Lagrangian. These are scalar loops which now have
massive propagators with thermal mass $\mu^2=\hat\Xi=e^2T^2/4$ from
(\ref{Xi}), from which the HTL result $\hat m_D^2$ has to be
subtracted as thermal counter-term. Indeed, without this subtraction,
the LO result would be generated a second time, since for
large loop momenta the thermal mass of the scalar is negligible.
Because of the subtraction, the one-loop integrals are now
receiving their leading contributions from soft loop momenta $k\sim eT$:
\bea\label{dPi00}
&&\delta\Pi_{00}(0,q) =
e^2 \sum_K {4 k_0^2\0 (K^2-\mu^2)[(K-Q)^2-\mu^2]}
-e^2 \sum _K {2\0K^2-\mu^2} - \hat m_D^2 \nn\\
&=& {e^2\0\pi^2} \int_0^\infty \!\! dk \,k^2 \left\{
    {n(\sqrt{k^2+\mu^2}) \0 \sqrt{k^2+\mu^2}} \left[ 1 +
    {k^2+\mu^2\0kq}\ln\left|{2k+q\02k-q}\right| \right] -(\mu\to0) \right\}
\eea
To obtain the leading
contribution, we can replace $n(E)\to T/E$ which gives
\be
\delta\Pi_{00}(0,q) =- {e^2T\0\pi^2} \int_0^\infty \! dk \,
\left\{ { \mu^2 \0 k^2+\mu^2 } + 1 - {k\0q} \ln
  \left| {2k+q\02k-q} \right| \right\} + O(e^2 q^2 \ln(T)).
\ee
The NLO correction to the Debye mass is now given by evaluating
this at $q=i\hat m_D$. Incidentially, the above integral is
$q$-independent, as can be seen from an integration by parts,
which finally gives
\be\label{SEDmD}
m_D^2=\hat m_D^2 - {e^2 T \mu \0 2\pi} = {e^2T^2\03}\left(
1-{3\04\pi}e \right).
\ee
Notice that the perturbative result at NLO involves a single
power of $e$ and so is {\em non-analytic} in $\alpha=e^2/(4\pi)$.

The above calculation can actually be simplified
by noting \cite{Arnold:1993rz}
that the only terms capable of producing odd powers
in $e$ are the $n=0$ terms in the sums over Matsubara frequencies in
(\ref{dPi00}). For $n\not=0$, one has $-k_0^2=(2\pi n)^2 T^2\gg \mu^2$
so that the thermal masses can be expanded out, leading to
powers of $e^2$ only. Keeping only the $n=0$ contributions,
the first integral in (\ref{dPi00}) vanishes, and we have
\be
\delta\Pi_{00}(0,q) = 2e^2 T \sigma^{2\epsilon}
\int {d^{3-2\epsilon}k\0(2\pi)^{3-2\epsilon}}{1\0k^2+\mu^2}
\stackrel{\epsilon\to0}{\longrightarrow} -{e^2T\mu\02\pi}
\ee
in agreement with (\ref{SEDmD}).
Here we have introduced dimensional regularization (with
mass scale $\sigma$) to render the
integral finite as in Ref.~\cite{Arnold:1993rz}.

As we shall see presently, this simplified resummation 
by {\em dimensional reduction} is only
possible in the static case ($q_0=0$), and not for
dynamical quantities. 

\subsubsection{Plasma frequency}

Propagating modes exist for frequencies $q_0\ge \omega_{\rm pl.}$,
and this plasma frequency is the same for the transverse and
longitudinal modes, because it corresponds to the long-wavelength
limit $q\to0$, and with $\1q=0$ there is no way to distinguish
the polarizations. In the HTL approximation, $\omega_{\rm pl.}^2=
\hat m_D^2/3=e^2T^2/9$.

The NLO correction in the case of scalar electrodynamics
can be calculated in full analogy to (\ref{dPi00}), but now
with $q_0=\omega_{pl.}$ and $\1q\to0$. Because of $\1q=0$,
the angular integrals are now trivial and one finds   
\be\label{dPil}
\delta\Pi_\ell(q_0,0)=\delta\Pi_t(q_0,0) 
=-{e^2T\02\pi}\left\{ \mu+{4\03q_0^2}
\left([\mu^2-q_0^2/4]^{3/2}-\mu^3\right) \right\}.
\ee
Evaluated at $q_0=\omega_{\rm pl.}$ this gives
$\delta\omega_{\rm pl.}^2/\omega_{\rm pl.}^2=-e(8\sqrt2-9)/(2\pi)
\approx -0.37 e$.
Notice that without resumming $\mu$, the result would have
been completely misleading: evaluating the unresummed result,
i.e.~(\ref{dPil}) with $\mu=0$, at $q_0=\omega_{\rm pl.}$ would
have given a purely imaginary result that one would have wrongly
identified with a damping constant.\footnote{In scalar
electrodynamics, unlike QCD, the plasmon damping is a higher
order effect because the scalar HTL quasiparticles are
heavier than plasmons and moreover do not have Landau damping
cuts in the HTL approximation.} 

Furthermore, the correct result (\ref{dPil}) is now only obtained
if the nonstatic modes are resummed along with the static ones.
If one  keeps only the zero modes and ignors that in the
imaginary time formalism the external frequency $q_0$
has to be a multiple of $2\pi i T$ (and so cannot be soft and
nonzero), but immediately `continues' to $q_0=\omega_{\rm pl.}\sim eT$,
one would find
\be\label{dPilstatic}
\delta\Pi_{\ell,t}(q_0,0)|_{\rm 0-mode \atop contr.} 
=-{e^2 T\06\pi}
\left\{{2\0q_0}\left[{\mu\0q_0}-\sqrt{{\mu^2\0q_0^2}-1}\right](q_0^2-\mu^2)
+\mu \right\}
\ee
which clearly differs from (\ref{dPil}). The resulting 
$\delta\omega_{\rm pl.}^2$ would in fact be only about a quarter of
the true result. So dynamic quantities require the full HTL
resummation method; resumming only the zero modes is not
sufficient (see also \cite{Elze:1988rh}).

\subsection{NLO corrections for QCD quasiparticles}
\label{NLOQCD}

The calculation of NLO corrections to the long-wavelength plasmons
in QCD was in fact one of the first applications of the HTL
resummation programme. In particular, the damping constant
of order $g^2T\sim g\omega_{\rm pl.}$ was the subject of
a long controversy (in particular with regard to its
gauge-fixing \hbox{(in)}\-dependence) 
before it was calculated in Ref.~\cite{Braaten:1990it}
with the gauge independent\footnote{In fact, later investigations
using covariant gauges encountered again gauge dependences
\cite{Baier:1992dy,Baier:1992mg}, 
which are removed, however, when an infrared cut-off
is retained while taking the on-mass-shell limit \cite{Rebhan:1992ak}.
This means that there is an infrared singularity in the
residue of the pole, which is gauge dependent, but the position
of the singularity (no longer a simple pole)
is gauge independent in accordance with
the theorems of Refs.~\cite{Kobes:1990xf,Kobes:1991dc}.}
result 
\be
\gamma\approx 0.264 \sqrt{N}\, g\, \omega_{\rm pl.}
\ee
(for pure-glue QCD). The analogous calculation for the damping constant of
long-wavelength fermionic quasiparticles was carried out
in Refs.~\cite{Kobes:1992ys,Braaten:1992gd}. The significance
of these results is that gluonic and fermionic quasiparticles,
which in the HTL approximation appear to be stable, experience
damping. For $g\ll 1$, they are weakly damped, whereas for $g\sim 1$
which is more relevant for experimentally accessible quark-gluon plasmas,
the damping is significant: $\gamma \sim \2\omega_{\rm pl.}$.

The NLO correction to the gluonic plasma frequency has also been
calculated
\cite{Schulz:1994gf} with the result 
$\delta\omega_{\rm pl.}^2/\omega_{\rm pl.}^2
\approx -0.18 \sqrt{N}\,g$.

The NLO correction to the Debye mass, however, runs into
IR problems. Naively one would expect problems from the
masslessness of magnetostatic gluons only at two-loop order
resummed perturbation theory, when their self-interactions
become relevant. However, because gauge independence requires
evaluation on mass-shell (which in the case of the Debye mass
means $q_0=0$, $\1q^2=-\hat m_D^2$), there appear `mass-shell
singularities' caused by the massless magnetostatic modes.
Because these singularities are only logarithmic, the leading
log is perturbatively calculable and reads
\cite{Rebhan:1993az,Rebhan:1994mx}
\be
\delta m_D^2/\hat m_D^2= {N\02\pi}\sqrt{6\02N+N_f} g \log{1\0g} + 
\mathcal O(g).
\ee
For small coupling, the logarithm dominates over the
non-perturbative constant behind the logarithm, and
thus the perturbative prediction is that of a {\em positive}
correction to the screening mass. Indeed, lattice simulations
of both (gauge-fixed) chromo-electrostatic propagators
\cite{Cucchieri:2001tw} and gauge-invariant lattice definitions
of the nonabelian Debye mass \cite{Arnold:1995bh,Laine:1999hh}
give significant positive corrections to the HTL value.\footnote{These
positive corrections are so large in fact that the nonperturbative
contributions at order $g$ dominate over the HTL result in the
gauge-invariant lattice definitions. When extracted from
lattice propagators, which also give gauge-independent results
\cite{Cucchieri:2001tw},
these nonperturbative contributions are considerably smaller
and about 1/3 of them can be accounted for by one-loop resummed
perturbation theory if one introduces a simple phenomenological
magnetic screening mass taken from the lattice
\cite{Rebhan:1994mx,Rebhan:2001wt}.}

Such a logarithmic sensitivity to the nonperturbative physics
of the chromo-magnetostatic sector has in fact been encountered
earlier in the damping of hard excitations
\cite{Pisarski:1989vd,Lebedev:1991un,Rebhan:1992ca}.
More generally, it arises for all propagating modes
\cite{Pisarski:1993rf} as well as for
all finite screening lengths \cite{Flechsig:1995sk}.
For nonzero wave-vector $\1q$, only the real corrections to
the dispersion law $\omega=\omega(q)$ turn out to be IR safe
in one-loop resummed perturbation theory.

\section{HTL-resummed thermodynamics}

In the previous section we have seen that dynamic quantities
cannot be treated by the simplified resummation scheme that
resums only static modes. For static quantities like the 
thermodynamic potential, however, such a resummation works
in the sense that it gives a scheme to systematically
compute the series expansion in powers (and log's) of the
coupling. This calculation has been performed to order $\alpha_s^{5/2}$
in QCD \cite{Arnold:1995eb} with the result (for pure glue)
\bea
&&P=-\Omega/V={8\pi^2 T^4\045} \biggl[ 1 - {15\alpha_s\04\pi} 
+ 30 ({\alpha\0\pi})^{3/2} + {135\02}({\alpha\0\pi})^2 \log {\alpha\0\pi}
\\
&-&\!{165\08}\left(\log{\bar\mu\02\pi T}-11.49 \right)({\alpha\0\pi})^2
+{495\02} \left(\log{\bar\mu\02\pi T}-3.23 \right)({\alpha\0\pi})^{5/2}
+\ldots 
\biggr].\nn
\eea
Unfortunately, this 
is very poorly convergent: only when $\alpha_s < 0.05$
one has apparent convergence, but this corresponds
to temperatures higher than $10^5 T_c$ !

In what follows we shall attempt
a different route that resums also the nonstatic modes,
and tries to keep resummation effects even when they are formally of
higher order than that achievable at a given loop order.

\subsection{Screened perturbation theory}

In scalar $\varphi^4$ theory, it has been shown 
\cite{Karsch:1997gj,Andersen:2000yj}
that the
convergence of thermal perturbation theory can be
improved if the thermal mass of the scalar quasiparticles
is kept within thermal integrals and not treated as
proportional to a coupling constant when setting up
the perturbation series. Technically, this is just as
in (\ref{resum}), but without the requirement that
the resummation has to take place in soft quantities only.
Because this changes the UV structure at any finite order
of perturbation theory, this introduces new UV divergences
and associated renormalization scheme dependences,
which in principle can become arbitrarily large. But
starting from two-loop order, these can be minimized
if the thermal mass used in screened perturbation theory
is determined by a variational principle, i.e.~a prinicple
of `minimal sensitivity' to the mass parameter
used in this reorganization of perturbation theory 
\cite{Andersen:2000yj,Rebhan:2000uc}.

In Refs.~\cite{Andersen:1999fw,Andersen:1999sf}, this approach
has been adapted to a one-loop calculation of the thermodynamic
potential of QCD where in place of a simple mass term the
gauge-invariant HTL effective action is used. While the
leading-order interaction term $\propto g^2$ is incomplete
(in fact, it is over-included), it does contain the plasmon
term $\propto g^3$ without leading to the disastrous result
of a thermodynamic pressure in excess of the Stefan-Boltzmann limit.
However, it remains to be seen if this is still true
in the (technically very difficult)
two-loop approximation, which presumably has to
contribute with positive sign to make up for the over-included
negative leading-order interaction term of the one-loop approximation.

In the following, I shall present instead the result of a different
strategy. Instead of using the HTL effective action as a
gauge invariant mass term for an optimization of
perturbation theory, which does not (have to) care whether
the HTL effective action remains accurate for hard momenta
(which it does not), the formalism of self-consistent 
``$\Phi$-derivable'' \cite{Baym:1962} approximations
will be invoked to find expressions that keep resummation
effects for both soft and hard momenta.
As we shall see, the leading-order effects will
arise exclusively from kinematical regimes where the
HTL approximation remains justifiable.
Moreover, it will be possible to avoid the spurious
UV problems of screened perturbation theory. 

\subsection{Approximately self-consistent resummations}

In the Luttinger-Ward representation of
the thermodynamic potential $\Omega=-PV$ \cite{Luttinger:1960} (to
particle physicists often more familiar as the composite operator effective
potential \cite{Cornwall:1974vz})
is expressed as a functional of full propagators $D$ and
two-particle irreducible diagrams. Considering for
simplicity a scalar field theory for the moment, $\Omega[D]$ has the form
\bea
\label{LW}
\Omega[D]&=&-T \log Z=\8\2 T \,\Tr \log D^{-1}-\8\2 T \,\Tr\, \Pi D
+T \Phi[D]\nn\\
&=&\int\!{d^4k\0(2\pi)^4}n(\omega) \Im \left[
\log D^{-1}(\omega,k)-\Pi(\omega,k) D(\omega,k) \right]+T\Phi[D],\;\;
\eea
where $\Tr$ denotes the trace in configuration space,
and
$\Phi[D]$ is the sum of the 2-particle-irreducible ``skeleton''
diagrams
\be\label{skeleton}
-\Phi[D]= 
\includegraphics[bb = 50 390 550 430,width=5cm]{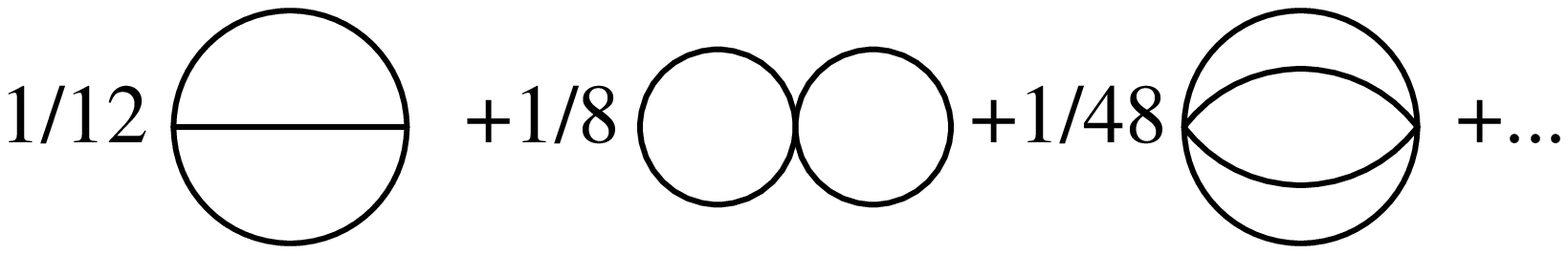}
\ee
The self energy $\Pi=D^{-1}-D^{-1}_0$, 
where $D_0$ is the bare propagator, is
related to $\Phi[D]$ by 
\be\label{PhiPi}
\delta\Phi[D]/\delta D=\8\2\Pi.
\label{Pi}
\ee
An important property of the functional  $\Omega[D]$, which is easily verified
using  (\ref{PhiPi}), is that it is stationary under variations of
$D$:
\be\label{selfcons}
{\delta\Omega[D] / \delta D}=0.
\ee
Self-consistent (``$\Phi$-derivable'')
\cite{Baym:1962} approximations
are  obtained by selecting a class of skeletons in 
$\Phi[D]$ and calculating $\Pi$ from Eq.~(\ref{Pi})
above, preserving the stationarity condition.

The stationarity of $\Omega[D]$ has an interesting consequence for  the
entropy ${\cal S}=-{\6(\Omega/V)/\6T}$.
Because of  
Eq.~(\ref{selfcons}), the temperature derivative of the spectral
density in the dressed propagator cancels out and only the explicit
Bose-Einstein factors need to be differentiated in (\ref{LW}),
yielding \cite{Riedel:1968,Vanderheyden:1998ph,Blaizot:1999ip,Blaizot:2000fc}
\bea\label{Ssc}
{\cal S}&=&-\int\!\!{d^4k\0(2\pi)^4}{\6n(\omega)\0\6T} 
\Im \log D^{-1}(\omega,k) \nn\\
&&+\int\!\!{d^4k\0(2\pi)^4}{\6n(\omega)\0\6T} \Im\Pi(\omega,k) \Re D(\omega,k)+{\cal S}'
\eea
with
\be
{\cal S}'=-{\6(T\Phi)\0\6T}\Big|_D+
\int\!\!{d^4k\0(2\pi)^4}{\6n(\omega)\0\6T} \Re\Pi\, \Im D=0
\ee
up to terms that are of loop-order 3 or higher.  Thus, in contrast to
$\Omega$, where
$\Phi$ contributes already to order
$g^2$ in perturbation theory, Eq.~(5) with ${\cal S}'=0$ is perturbatively
correct to order $g^3$.  The
first two terms in Eq.~(\ref{Ssc}) represent essentially the entropy of
``independent quasiparticles'', while ${\cal S}'$ may  be viewed as the
residual interactions among these quasiparticles \cite{Vanderheyden:1998ph}.

The same simplification holds true in the presence of fermions and,
with nonzero chemical potential, extends to the fermion
density 
$\mathcal N=-{\6(\Omega/V)/\6\mu}$ \cite{Blaizot:1999ap,Blaizot:2000fc}.
In a {\em self-consistent} two-loop approximation one thus has
the remarkably simple formulae, now for general theories
\bea
\label{S2loop}
{\cal S}\!&=&\!-\tr \int{d^4k\0(2\pi)^4}{\6n(\omega)\0\6T} \left[ \Im 
\log D^{-1}(\omega,k)-\Im \Pi(\omega,k) \Re D(\omega,k) \right] \nn\\
&&-2\,\tr \int{d^4k\0(2\pi)^4}{\6f(\omega)\0\6T} \left[ \Im
\log S^{-1}(\omega,k)-\Im \Sigma(\omega,k) \Re S(\omega,k) \right],\;\;\;\; \\
\label{N2loop}
{\cal N}\!&=&\!-2\,\tr \int{d^4k\0(2\pi)^4}{\6f(\omega)\0\6\mu} \left[ \Im
\log S^{-1}(\omega,k)-\Im \Sigma(\omega,k) \Re S(\omega,k) \right],\;\;\;\;
\eea
where $n(\omega)=(e^{\beta\omega}-1)^{-1}$,
$f(\omega)=(e^{\beta(\omega-\mu)}+1)^{-1}$, and 
``tr'' refers to all discrete labels,
including spin, color and flavor when applicable.

In gauge theories, the above expressions have to be augmented by
Fad\-deev-Popov ghost contributions which enter like bosonic fields
but with opposite over-all sign, unless a gauge is used where
the ghosts do not propagate such as in axial gauges.
But because $\Phi$-derivable approximations do not generally
respect gauge invariance,\footnote{For this, one would have
to treat vertices on an equal footing with self-energies, which
is in principle possible using the formalism
developed in Ref.~\cite{Freedman:1977xs}.}
the self-consistent two-loop approximation
will not be gauge-fixing independent.
It is in fact not even clear that the corresponding gap equations
(\ref{Pi}) have solutions at all or that one can renormalize
these (nonperturbative) equations.

For this reason, we shall construct {\em approximately self-consistent}
solutions which are gauge invariant and which
maintain equivalence with conventional
perturbation theory up to and including order $g^3$,
the maximum (perturbative) accuracy of the two-loop approximation
for $\Phi$. For these approximations it will be sufficient
to keep only the two transverse structure functions of the
gluon propagator and to neglect ghosts.

For soft momenta, we know that the leading order contribution
is given by the HTL's, and indeed there is no HTL ghost self-energy.
For hard momenta, one can identify the contributions to
(\ref{S2loop}) below
order $g^4$ as those linear in the self-energies,
\bea\label{Shard}
\mathcal S^{\rm hard}&=&\mathcal S_0
+2N_g\int\!\!{d^4k\0(2\pi)^4}\,{\6n\0\6T}\,\Re{\Pi_t}\,
\Im\frac{1}{\omega^2-k^2}\nn\\
&-&\!\!4NN_f\int\!\!{d^4k\0(2\pi)^4}\,{\6f\0\6T}\,\Bigl\{
\Re\Sigma_+\Im\frac{-1}{\omega-k}\,-\,
\Re\Sigma_-\Im\frac{-1}{\omega+k}\Bigr\}
\eea
considering now a gauge theory with $N_g$ gluons
and $N_f$ fermion flavors.
Because the imaginary parts of the free propagators restrict their
contribution to the light-cone, only the light-cone projections
of the self-energies
enter. At order $g^2$ this is exactly 
given by the HTL results, without having to assume soft $\omega,k$
\cite{Kraemmer:1990drA,Flechsig:1996ju}
(as in the above example of scalar electrodynamics,
see Eqs.~(\ref{PimumuSED}), (\ref{Xi}))
\bea\label{mas2}
&&\Re\Pi_t^{(2)}(\omega^2=k^2)=\hat\Pi_t(\omega^2=k^2)=
\8\2\hat m_D^2\equiv m_\infty^2,\\
&&2k \,\Re\Sigma_\pm^{(2)}(\omega=\pm k)=2k \,\hat\Sigma_\pm(\omega=\pm k)=
2 \hat M^2 \equiv M_\infty^2,\label{Mas2}
\eea
and without contributions from the other components of $\Pi_{\mu\nu}$
and the Faddeev-Popov self-energy.

There is no contribution $\propto g^2$ from soft momenta
in (\ref{S2loop}) and (\ref{N2loop}) so that one is left with
remarkably simple general formulae for the leading-order interaction
contributions to the thermodynamic potentials expressed
through the asymptotic thermal masses of the bosonic and
fermionic quasiparticles:
\be\label{SNBF2}
{\cal S}^{(2)} =-  T\left\{\sum_B { m_{\infty\,B}^2 \0 12}\,+\,
\sum_F { M_{\infty\,F}^2\0 24}\right\},\quad
{\cal N}^{(2)}=-\,{1\0 8\pi^2}\sum_F \mu_F M_{\infty\,F}^2.\ee
Here the sums run over all the bosonic ($B$) and fermionic ($F$)
degrees of freedom (e.g. 4 for each Dirac fermion), 
which are allowed to have different asymptotic
masses and, in the case of fermions, different chemical potentials.

Turning now to the next order, $g^3$, let us first recapitulate
how this ususally arises in the thermodynamic potential.
Since this is a static quantity, we can use the imaginary-time
formalism and concentrate on the zero-modes as in Sect.~\ref{mDSED}.
For $\omega=0$ the leading contributions to the self-energies
are $\hat\Pi_\ell(0,k)=\hat m_D^2\propto (gT)^2$ and $\hat\Pi_t(0,k)=0$.
This gives the ``plasmon-effect''\footnote{Obviously a misnomer.
It is caused exclusively by the Debye screening mass, not the
plasmon mass = plasma frequency $\not= \hat m_D$.} term
\be\label{P3}
P^{(3)}=-N_g T \int\!\!{d^3 k\0(2\pi)^3}\left[\log\left(
1+\frac{\hat  m_D^2}{k^2}\right)-\frac{\hat  m_D^2}{k^2}\right]
\,=\,N_g \frac{\hat m_D^3 T}{12\pi}.
\ee

In deriving the self-consistent expressions for entropy and density,
Eqs.\ (\ref{S2loop}) and (\ref{N2loop}), we can no longer use
this argument to extract the order $g^3$ term, for we have
first rewritten the thermodynamic potential in terms of
real-time propagators and self-energies and then used
stationarity to drop $T$ and $\mu$ derivatives of the spectral
densities hidden in the full propagators. In fact,
from the second line of (\ref{LW}), one can still isolate
the zero-mode contribution (\ref{P3}) by
$\int {d\omega\0\pi\omega}\Im[\dots]=[\ldots](\omega=0)$,
but in (\ref{S2loop}) and (\ref{N2loop}) we have products
of imaginary and real parts times statistical distribution functions.

Indeed, the order $g^3$ contributions now arise from both soft and
hard momentum regimes. In ${\cal S}^{\rm soft}$ such
contributions are due to the singular behaviour of
$\6 n/\6T \sim 1/\omega$ which does not allow us to
expand out the self-energy insertions perturbatively,
but on dimensional grounds gives a contribution $\sim \Pi^{3/2}$.
In ${\cal S}^{\rm hard}$, on the other hand, where we could
expand out the self-energies as in Eq.~(\ref{Shard}), 
$g^3$ contributions arise from
NLO order contributions to $\Pi$ and $\Sigma$ themselves.
(In the case of the pressure, this did not happen because
of the stationarity property of the pressure \cite{Blaizot:2000fc}.)

With this insight, we can now formulate an {\em approximately
self-consistent} dressing of the propagators that is in line
with the maximum perturbative accuracy of the $\Phi$-derivable two-loop
approximation: for soft momenta, we take the (gauge-invariant
and gauge-independent)
HTL expressions for self-energies and propagators; for
hard momenta, where according to (\ref{Shard}) only the
light-cone limit of the self-energies contribute below
order $g^4$, the correct leading-order contribution is
still given by the HTL expressions (\ref{mas2}), (\ref{Mas2}),
but in order to include the $g^3$ contributions completely,
we also require the NLO corrections to the
on-light-cone self-energies. The latter can be
calculated by standard HTL perturbation theory, and
the theorems of Ref.~\cite{Kobes:1991dc} ensure their
gauge independence.

\subsubsection{HTL approximation}

As a first approximation let us consider one which only uses
the HTL expressions without NLO corrections thereof.
We have seen that this gives the correct leading-order
interaction term $\propto g^2$, some part of the $g^3$
contribution, and infinitely many formally higher-order
terms as well, since we are going to use (\ref{S2loop})
and (\ref{N2loop}) ``non-perturbatively'', i.e.~without
expanding out in powers of $g$ and truncating.
We can do so because expressions (\ref{S2loop})
and (\ref{N2loop}) are manifestly UV finite, for
they involve only the derivatives of the statistical distribution
functions---in (\ref{LW}), the $T=0$ UV-divergences
are contained in the integration domain $\omega\to-\infty$,
where the undifferentiated $n$'s and $f$'s do not fall off
exponentially.

Using the HTL expressions in (\ref{S2loop})
and considering for simplicity the pure-glue case, we obtain
two physically distinct contributions. The first 
corresponds to the transverse
and longitudinal gluonic quasiparticle poles,
\be\label{SQP}
{\cal S}^{\rm QP}_{\rm HTL}=- N_g\int\!\!{k^2\,dk\02\pi^2} {\6\0\6T}
\Bigl[2T\log(1-e^{-\omega_t(k) / T}) 
+ T\log{1-e^{-\omega_\ell(k) / T}\01-e^{-k/T}} \Bigr],
\ee
where only the explicit $T$ dependences are to be differentiated,
and not those implicit in the HTL dispersion laws $\omega_t(k)$ and 
$\omega_\ell(k)$. Secondly, 
there are the Landau-damping contributions which read
\bea\label{SLD}
{\cal S}^{\rm LD}_{\rm HTL}&=&
-N_g \int\limits_0^\infty{ k^2 dk \0 2\pi^3} 
\int\limits_0^k \! d\omega {\6n(\omega)\0\6T} 
\Bigl\{ 2 \arg [ k^2-\omega^2+\hat \Pi_t ] \nn\\
&&\hspace{-2.2cm}-2\,\Im \hat \Pi_t \,\Re[\omega^2-k^2-\hat \Pi_t]^{-1} 
+ \arg [ k^2+\hat \Pi_\ell ] - \Im \hat \Pi_\ell \,\Re[k^2+\hat \Pi_\ell]^{-1} 
\Bigr\}.
\eea
The usual perturbative $g^2$-contribution (\ref{SNBF2}) is contained in the 
first term of Eq.~(\ref{SQP}); 
all the other terms in Eqs.~(\ref{SQP}),(\ref{SLD})
are of order $g^3$ in a small-$g$ expansion.

In the HTL approximation, only the soft plasmon effect $\sim g^3$
contained
in $\mathcal S^{\rm soft}$ is present, which turns out to
equal
\be
{\cal S}^{\rm soft}_{3}={\6 P^{(3)}\0\6 T}\Big|_{\hat m_D},
\ee
which is $\frac14 {\cal S}_3$ in the case of pure glue.
However, this identification requires a peculiar sum rule
\bea\label{DELTASS}
\Delta {\cal S}_3&\equiv &
N_g\int\!\!{d^4k\0(2\pi)^4}\,{1\0\omega}\,
\biggl\{2\,\Im\hat \Pi_t \Re \Bigl(\hat  D_t - D_t^{(0)}\Bigr)
-\Im\hat \Pi_\ell \Re\Bigl(\hat  D_\ell-D_\ell^{(0)}\Bigr)\biggr\}\nn\\
&&\equiv \Delta {\cal S}_t^{(3)}+\Delta {\cal S}_\ell^{(3)}=0\,,\eea
which we found to hold numerically by cancellations in more
than 8 significant digits. Rather unusually (cp.~Ref.~\cite{LeB:TFT}), this
does not hold separately for the longitudinal and transverse
sector and moreover holds only after carrying out both, the frequency
and the momentum integrations in (\ref{DELTASS}).

Although the $g^3$ term in $\mathcal S_{\rm HTL}$ is only a fraction
of the full one, it would make similar troubles when expanded out
perturbatively, throwing away all higher-order terms in $g$:
for large enough $g$, the perturbative approximation would lead
to an entropy in excess of the Stefan-Boltzmann limit. $\mathcal S_{\rm HTL}$,
on the other hand, is a monotonically decreasing function of
$\hat m_D/T \sim g$.

\subsubsection{Next-to-leading approximation}

The plasmon term $\sim g^3$ 
becomes complete only upon inclusion
of the next-to-leading correction to the asymptotic thermal
masses $m_\infty$ and $M_\infty$. These are determined
in standard HTL perturbation theory through
\be
\begin{array}{l}
\delta m_\infty^2(k)=\Re \delta\Pi_T(\omega=k) \\
=\Re(\begin{picture}(0,0)(0,0)
\put(25,0){\small +}
\put(56,0){\small +}
\put(104,0){\small +}
\put(157,0){$|_{\omega=k}$}
\end{picture}
\!\!\includegraphics[bb=145 430 500 475,width=5.5cm]{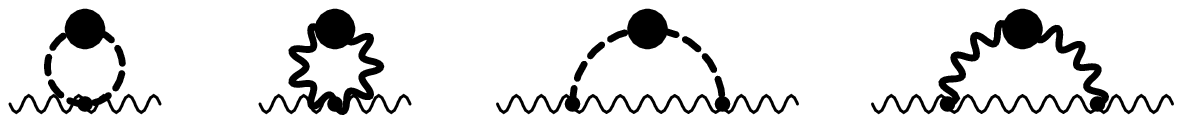})
\end{array}
\ee
where thick dashed and wiggly lines with a blob represent
HTL propagators for longitudinal and transverse polarizations, respectively.
Similarly,
\be
{1\02k}\delta M_\infty^2(k)=\delta\Sigma_+(\omega=k) 
=\Re(\begin{picture}(0,0)(0,0)
\put(42,0){\small +}
\end{picture}
\includegraphics[bb=75 430 285 475,width=3.2cm]{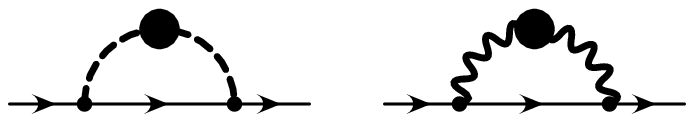})
|_{\omega=k}\;.
\ee
The explicit proof that these contributions indeed restore the
correct plasmon term is given in Ref.~\cite{Blaizot:2000fc}.

These corrections to the asymptotic thermal masses are, in contrast
to the latter, nontrivial functions of the momentum, which can
be evaluated only numerically. However, as far as the generation
of the plasmon term is concerned, these functions contribute
in the averaged form
\be\label{deltamasav}
\bar\delta m_\infty^2={\int dk\,k\,n_{\rm BE}'(k) \Re \delta\Pi_T(\omega=k) 
\0 \int dk\,k\,n_{\rm BE}'(k)}
\ee
(cp.~Eq.~(\ref{SNBF2})) and similarly
\be\label{deltaMasav}
\bar\delta M_\infty^2={\int dk\,k\,n_{\rm FD}'(k) \Re 
2k \delta\Sigma_+(\omega=k) 
\0 \int dk\,k\,n_{\rm FD}'(k)}\;.
\ee
These averaged asymptotic thermal masses turn out to be given
by the remarkably simple expressions \cite{Blaizot:2000fc}
\be
\label{deltamas}
\bar\delta m_\infty^2=-{1\02\pi}g^2NT\hat m_D,\quad
\bar\delta M_\infty^2=-{1\02\pi}g^2C_fT\hat m_D,
\ee
where $C_f=N_g/(2N)$. Since the integrals in
(\ref{deltamasav}) and (\ref{deltaMasav}) are dominated by hard
momenta, these thermal mass corrections only pertain to hard
excitations. Indeed, in Sect.~\ref{NLOQCD} we have seen
that e.g. the plasmon mass at $k=0$ receives a different,
namely smaller correction, whereas the NLO contribution to the
Debye mass is even positive and logarithmically enhanced. 

Pending the full evaluation of the NLO corrections to $\Re\delta\Pi$
and $\Re\delta\Sigma$, it has therefore been proposed
in Refs.~\cite{Blaizot:1999ip,Blaizot:1999ap,Blaizot:2000fc} 
to define a next-to-leading approximation through (for gluons)

\be
{\cal S}_{NLA}={\cal S}_{HTL}\Big|_{\rm soft}+
{\cal S}_{\bar m_\infty^2}\Big|_{\rm hard},
\ee
where $\bar m_\infty^2$ includes (\ref{deltamas}).
To separate soft ($k\sim \hat m_D$) and hard ($k \sim 2\pi T$) momentum
scales, we introduce the intermediate scale
$\Lambda=\sqrt{2\pi T\hat m_D c_\Lambda}$
and consider a variation of $c_\Lambda=\2\ldots 2$ as part of
our theoretical uncertainty.

Another crucial issue concerns the definition of the corrected
asymptotic mass $\bar m_\infty$. For the range of coupling constants
of interest ($g\gtrsim 1$), the correction $ |\bar\delta m_\infty^2|$
is greater than the LO value $m_\infty^2$, leading to tachyonic
masses if included in a strictly perturbative manner.

However, this problem is not at all specific to QCD. In the
simple $g^2\varphi^4$ model, one-loop resummed perturbation theory
gives 
\be\label{mstrpert}
m^2=g^2T^2(1-{3\0\pi}g)
\ee 
which also turns tachyonic
for $g\gtrsim 1$. On the other hand, the self-consistent
solution of the corresponding
one-loop gap equation \cite{Dolan:1974qd,Drummond:1997cw}
\be
m^2=\Pi(m)=12g^2\sum_K{-1\0K^2-m^2}
\ee
(properly renormalized), whose perturbative expansion
begins exactly like (\ref{mstrpert}),
is a monotonic function in $g$. In fact,
it turns out that the first two terms in a $(m/T)$-expansion of
this gap equation,
\be\label{mtruncgap}
m^2=g^2T^2-{3\0\pi}g^2Tm,
\ee
which is perturbatively equivalent to (\ref{mstrpert}), has
a solution that is extremely close to that of the full gap
equation (for $\overline{\mbox{MS}}$ renormalization scales 
$\bar\mu \approx 2\pi T$) \cite{Blaizot:2000fc}.

In QCD, the (non-local) gap equations are way too complicated to
be attacked directly. We instead consider perturbatively equivalent
expressions for the corrected $\bar m_\infty$ which are monotonic
functions in $g$. Besides the solution to a quadratic equation
analogous to (\ref{mtruncgap}) we have tried the
simplest Pad\'e approximant $m^2=g^2T^2/(1+{3\0\pi}g)$, which
also gives a greatly improved approximation to the solution of
scalar gap equations. In QCD, our final results do not depend
too much on whether we use the Pad\'e approximant
\cite{Blaizot:1999ip,Blaizot:1999ap}
or a quadratic gap equation
\cite{Blaizot:2000fc}.

The main uncertainty rather comes from the choice of the
renormalization scale which determines the magnitude of
the strong coupling constant when this is taken as determined
by the renormalization group equation (2-loop in the following).

\begin{figure}[tb]
\includegraphics[bb=70 180 540 560,width=7cm]{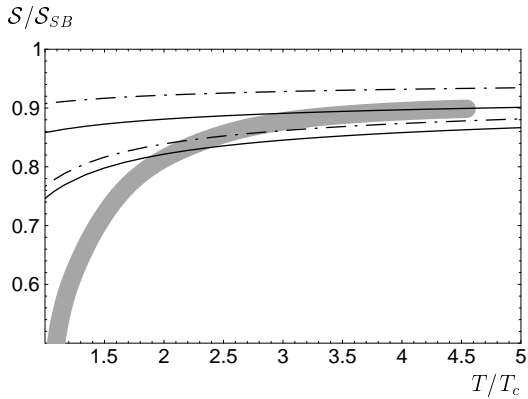}
\caption{Comparison of the lattice data for
the entropy of pure-glue SU(3) gauge theory of Ref.~\cite{Boyd:1996bx}
(gray band) with the range of ${\cal S}_{HTL}$ (solid lines)
and ${\cal S}_{NLA}$ (dash-dotted lines) for $\bar\mu=
\pi T\ldots 4\pi T$ and $c_\Lambda= 1/2 \ldots 2$.}\label{figSg}
\end{figure}

In Fig.~\ref{figSg}, the numerical results for the HTL entropy
and the NLA one are given as a function of $T/T_c$ with $T_c$
chosen as $T_c=1.14\Lambda_{\overline{\mathrm MS}}$. The full lines
show the range of results for
${\cal S}_{HTL}$ when the renormalization scale $\bar\mu$
is varied from $\pi T$ to $4\pi T$; the dash-dotted lines mark
the corresponding results for ${\cal S}_{NLA}$ with the
additional variation of $c_\Lambda$ from $1/2$ to 2. The
dark-gray band are lattice data from Ref.~\cite{Boyd:1996bx}.
Evidently, there is very good agreement for $T\gtrsim 2.5T_c$.

{}From the above results for the entropy density, one can
recover the thermodynamic pressure by simple integration,
$P(T)-P(T_0)=\int_{T_0}^T dT' S(T')$.
The integration constant $P(T_0)$, however, is a
strictly nonperturbative input. It cannot be fixed by
requiring $P(T=0)=0$, as this is in the confinement regime.
It is also not sufficient to know that $\lim_{T\to\infty}P=P_{\rm free}$
by asymptotic freedom. In fact, the undetermined integration constant
in $P(T)/P_{\rm free}(T)$ when expressed as a function
of $\alpha_s(T)$ corresponds to a term \cite{Blaizot:1999ap}
$C\exp\{-\alpha_s^{-1}
[4\beta_0^{-1}+O(\alpha_s)]\}$,
which vanishes for $\alpha_s\to 0$ with all derivatives
and thus is not fixed by any order of perturbation theory.
It is, in essence, the nonperturbative bag constant, which
can be added on to standard perturbative results, too. However,
in $P(T)/P_{\rm free}(T)$ this term becomes rapidly unimportant
as the temperature is increased, as it decays like $T^{-4}$.
Fixing it by $P(T_c)=0$, which is a good approximation in particular
for the pure-glue case because glue balls are rather heavy,
one finds again good agreement with lattice data for $T\gtrsim 2.5 T_c$.

This approach can be generalized \cite{Blaizot:1999ap,Blaizot:2000fc}
also to nonzero chemical potentials
$\mu_f$, where lattice data are not available\footnote{Lattice results
exist however for the response of thermodynamic quantities to
infinitesimal chemical potentials, namely for quark number
susceptibilities \cite{Gottlieb:1987ac,Gavai:2001fr,Gavai:2001ie},
and the above approach has been applied to those, too, by now
\cite{Blaizot:2001vr}.}. Simpler quasiparticle models 
\cite{Peshier:1996ty,Levai:1997yx}
have already been used to extrapolate lattice data to finite
chemical potential \cite{Peshier:1999ww}. 
The HTL approach offers a possible refinement,
but that has still to be worked out.

\section{Conclusion}

Hard thermal loops, the leading-order contributions to self-energies
and vertices at high temperature and/or density, form a gauge-invariant
basis for a systematic perturbative expansion, as long as one
does not run into the perturbative barrier formed by the
inherently nonperturbative sector of self-interacting
chromomagnetostatic modes. But in QCD one faces the
additional problem that corrections to leading-order results
are so large for almost all values of the coupling of interest
that they lead to a complete loss of (apparent) convergence.
However, we have seen that further resummations which keep
as much as possible of the effects of HTL resummation
without expanding in a power series in the coupling
may lead to results that remain valid down to temperatures
a few times the deconfinement phase transition temperature. 


\end{document}